\documentclass[aps,prl,twocolumn,showpacs,floatfix]{revtex4}
\usepackage{graphicx}
\usepackage{bm}
\usepackage{color}
\begin{document}

\title{Controlled Confinement of Half-metallic 2D Electron Gas in BaTiO$_3$/Ba$_{2}$FeReO$_6$/BaTiO$_3$ 
Heterostructures: A First-principles Study}

\author{Santu Baidya$^{1,\ast}$, Umesh V. Waghmare$^{1}$, Arun Paramekanti$^{2,3}$, Tanusri Saha-Dasgupta$^{4,\dagger}$}
\affiliation{$^{1}$ Jawaharlal Nehru Centre for Advanced Scientific Research,  Jakkur, Bangalore 560064, India. \\
$^{2}$ Department of Physics, University of Toronto, Toronto, Ontario, Canada M5S 1A7. \\
$^{3}$ Canadian Institute for Advanced Research, Toronto, Ontario, M5G 1Z8, Canada. \\
$^{4}$ Department of Condensed Matter Physics and Materials Science, S. N. Bose National Centre for Basic Sciences, Kolkata 700 098, India.}

\date{\today}

\begin{abstract}
Using density functional theory calculations, we establish that the half-metallicity of bulk Ba$_2$FeReO$_6$ 
survives down to 1 nm thickness in BaTiO$_3$/Ba$_{2}$FeReO$_6$/BaTiO$_3$ 
heterostructures grown along the (001) and (111) directions.
The confinement of the two-dimensional (2D) electron gas in this quantum well structure
arises from the suppressed hybridization between Re/Fe $d$ states and unoccupied Ti $d$ states,
and it is further strengthened by polar fields for the (111) direction. This mechanism, distinct from the 
polar catastrophe, leads to an order of magnitude stronger confinement of 
the 2D electron gas than that at the LaAlO$_3$/SrTiO$_3$ interface.
We further show low-energy bands of (111) heterostructure display nontrivial topological character.
Our work opens up the possibility of realizing ultra-thin spintronic devices.
\end{abstract}

\maketitle

A fundamental limit on device miniaturization is set by the size
at which the material properties change qualitatively.
This technologically important issue, has been studied for
ferroelectrics, both, theoretically \cite{nicola} and 
experimentally,\cite{fong} yielding a critical thickness of $\approx \! 12$\AA.\cite{fong}
For spintronic devices, it is similarly important to explore how thin one can make 
half-metals (HMs), i.e., materials which conduct in one spin channel and insulate in another. 
It is of great technological interest if such device can further be made operative
at high temperature.

In this Letter, we use first-principles density functional theory (DFT) to address this issue considering quantum 
wells of a double perovskite (DP) \cite{review1,tsd}  embedded in a wide band gap insulating oxide. 
In a heterostructure quantum well geometry, the electrons of the DP may be confined to 2D if
potential energy mismatch can be created between the transition metal (TM) ions in the DP and in the 
insulating oxide, as in case of ultra-thin metallic layers of SrTi$_{0.8}$Nb$_{0.2}$O$_{3}$ embedded in 
insulating SrTiO$_{3}$ \cite{ohta} or the (001) interface of SrTiO$_3$ with SrRuO$_3$.\cite{sro-sto}
This mechanism is different from polar catastrophe driven confinement of two dimensional electron gas, as found in well studied 
case of [100] interface of LaAlO$_3$ (LAO) and  SrTiO$_3$ (STO),\cite{nature427.423, nature419.378} or 
recently studied interfaces and quantum well structures formed with GdTiO$_3$ (GTO) or SmTiO$_3$ (SmTO) and STO.\cite{gto,smo}

In particular, we focus on quantum wells formed by embedding the DP Ba$_2$FeReO$_6$ (BFRO), in
the band insulator BaTiO$_3$ (BTO). The choice of BFRO is motivated by the fact that the bulk material is 
reported to be a half-metallic ferromagnet at room temperature, with T$_c$ $\approx$ 304K.\cite{PhysRevLett.98.017204} 
BTO is well-known to be a ferroelectric. However, we found that the ferroelectricity has
little effect on the half-metallic properties of the embedded BFRO quantum well. Thus, we only present results
here for the cubic paraelectric phase of BTO. \textcolor{black}{The comparison of structure
and electronic properties considering the polar BTO surface in its ferroelectric phase is given 
in the supplementary material (SM).} This choice of materials also ensures good lattice matching; the 
cubic lattice constants of BFRO and BTO being 4.025 \AA and 4.009 \AA, respectively. Finally, the 
synthesis of La$_{0.7}$Ca$_{0.3}$MnO$_3$ in contact with [001] surface of BaTiO$_3$ has been reported
in the literature,\cite{lee} giving us confidence in realization of our proposed system.
While oxide superlattices studied so far are grown along (001) direction, there has been recent progress on
growth along the (111) direction,\cite{Gibert_NatMat2012,Gray_APL2010,takagi2015} propelled, in part,
by theoretical predictions of exotic topological 
phases.\cite{Xiao2011,Fiete2012,Cook2014} \textcolor{black}{In particular, epitaxial growth of NdNiO$_3$ film on (111) 
direction has been achieved on a variety of substrate which includes LSAT, NdGaO$_3$ and LaAlO$_3$.\cite{catalano}}
We therefore consider two growth directions, (111) and (001). BFRO has a nominal oxidation state 
of 5+ for Re and 3+ for Fe in BFRO, as shown by x-ray magnetic circular dichroism (XMCD) experiments,\cite{PhysRevLett.98.017204} 
while Ti in BTO has a nominal oxidation state 4+. This features 
nonpolar interfaces with a cubic symmetry for (001) heterostructure [cf. Fig 1(b,d)], and two 
polar interfaces, one of $n$-type and another of $p$-type, with a hexagonal symmetry [cf. Fig. 1(a,c)] for 
(111) heterostructure. 

The 2D confinement in the BTO/BFRO/BTO quantum well structures studied here
provides significant improvement over that in LAO/STO in terms of, (i) 2D confinement length being an order of 
magnitude smaller (ii) complete spin-polarization of the 2D electron gas (2DEG) (iii) polarity control of the 2DEG,
suggestive of magneto-electric coupling, and (iv) realization of ultra-thin half metals with topological bands.
\textcolor{black}{We note here even the most recent report of 2DEG at the 
surface of STO\cite{sto} is found to be much less confined compared to that 
achieved in present study.}

\begin{figure}[t]
\includegraphics[width=8.6cm,keepaspectratio]{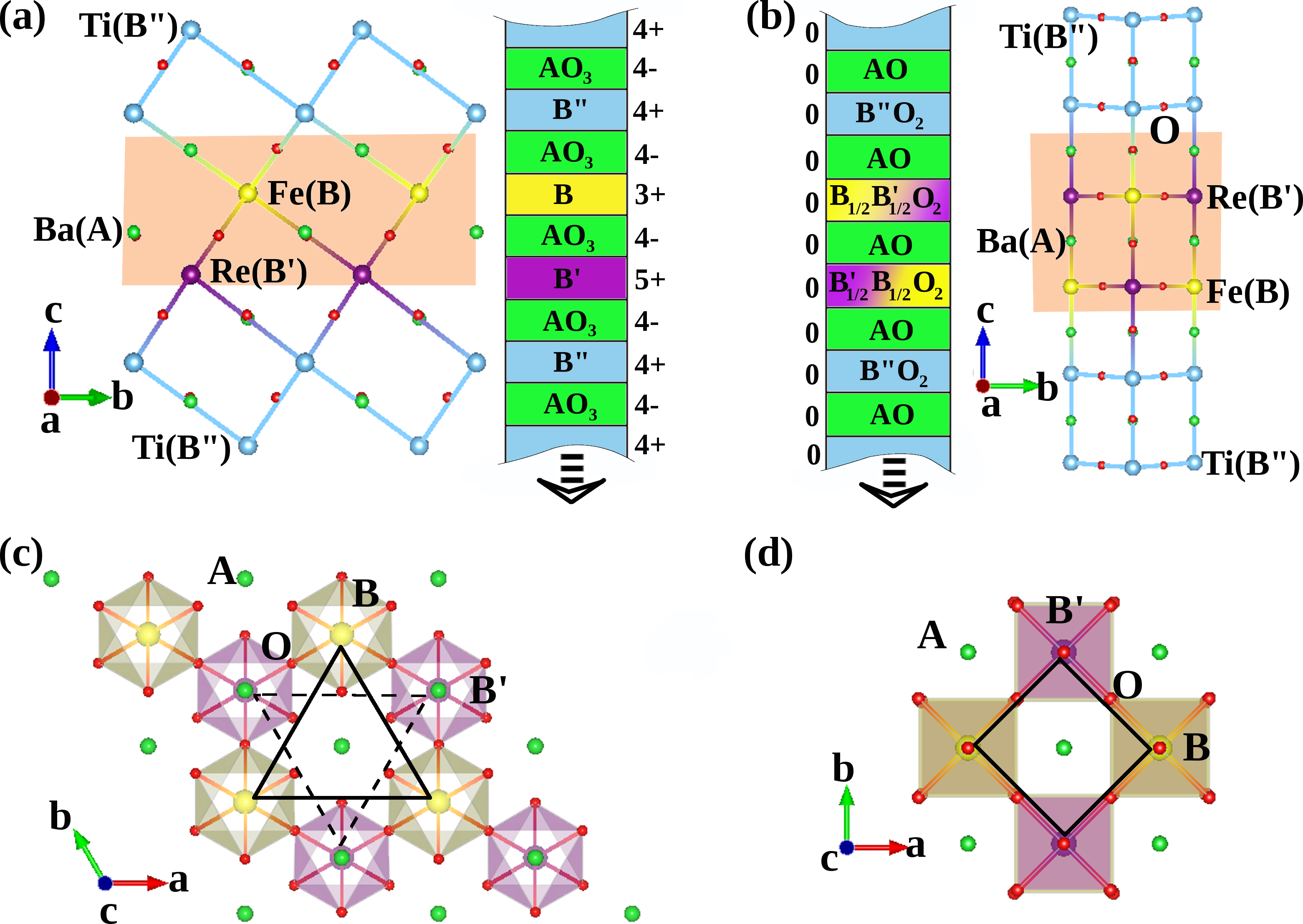}
\caption{(Color online) Geometry of BTO/BFRO/BTO heterostructures with c-axis oriented along the growth directions, (111)
[panels (a), (c)] and (001) [panels (b), (d)]. Layer-wise nominal charges for the two heterostructures are shown in 
panels (a) and (b). The inplane honeycomb lattice with two 
triangular sublattices on different layers in case of (111) heterostructure and the inplane square lattice in case of (001) heterostructure 
are marked in panels (c) and (d), respectively.}
\end{figure}

We use DFT calculations, within the generalized gradient approximation (GGA)\cite{ggapbe} for the exchange-correlation functional,
to study the electronic structure of bulk BFRO as well as the heterostructures. Our calculations are carried out using 
two choices of basis sets: (a) plane wave based pseudopotential method as implemented in the Vienna An-initio Simulation Package 
(VASP),\cite{vasp}
and (b) muffin tin orbital (MTO) based linear muffin tin orbital (LMTO)\cite{lmto} and N-th order MTO (NMTO) method.\cite{nmto}
The calculations include spin-orbit coupling (SOC), which is important for Re 5$d$ states. 
Details can be found in the supplementary material (SM).\cite{suppl}

Bulk Ba$_{2}$FeReO$_{6}$ contains rocksalt arrangement of Fe and Re which are connected via O along all three 
crystallographic directions.\cite{PhysRevLett.98.017204} The quantum well structures are simulated by 
considering a supercell of cubic BaTiO$_{3}$ along desired crystallographic direction, (111) or (001), and replacing Ti cations by Fe and Re cations 
in two adjacent middle layers along c-direction of supercell. 
%%[111\} oriented supercell consists of twelve BaO$_{3}$ layers, ten Ti layers, and 
%%two Fe and one Re layer. [001\} oriented supercell possesses eleven TiO$_2$ layers, nine BaO layers, and two Fe$_{0.5}$Re$_{0.5}$O$_{2}$ layers. 
The inplane lattice constants of the supercells are fixed to that of the BaTiO$_3$, which for (111) growth 
direction is $\sqrt{2/3}a_0$ and for (001) growth direction is $\sqrt{2}a_0$, $a_{0}$ being the cubic lattice constant of BTO. 
We completely relax the lattice parameter along the growth direction as well as
all the atomic positions. In the buckled honeycomb lattice geometry for the (111) growth direction, this introduces trigonal distortion in 
BO$_6$ octahedra, signaled by the deviation of $\angle$O-B-O from 90$^{o}$. The optimized (111) heterostructure shows this 
deviation to be 0.5$^{o}$ and 2.5$^{o}$ for ReO$_6$ and FeO$_6$ octahedra respectively. For the (001) heterostructure, the FeO$_6$ and ReO$_6$ octahedra 
upon optimization are found to be slightly elongated and contracted along the c-axis, respectively.

\begin{figure}
\includegraphics[width=8.4cm,keepaspectratio]{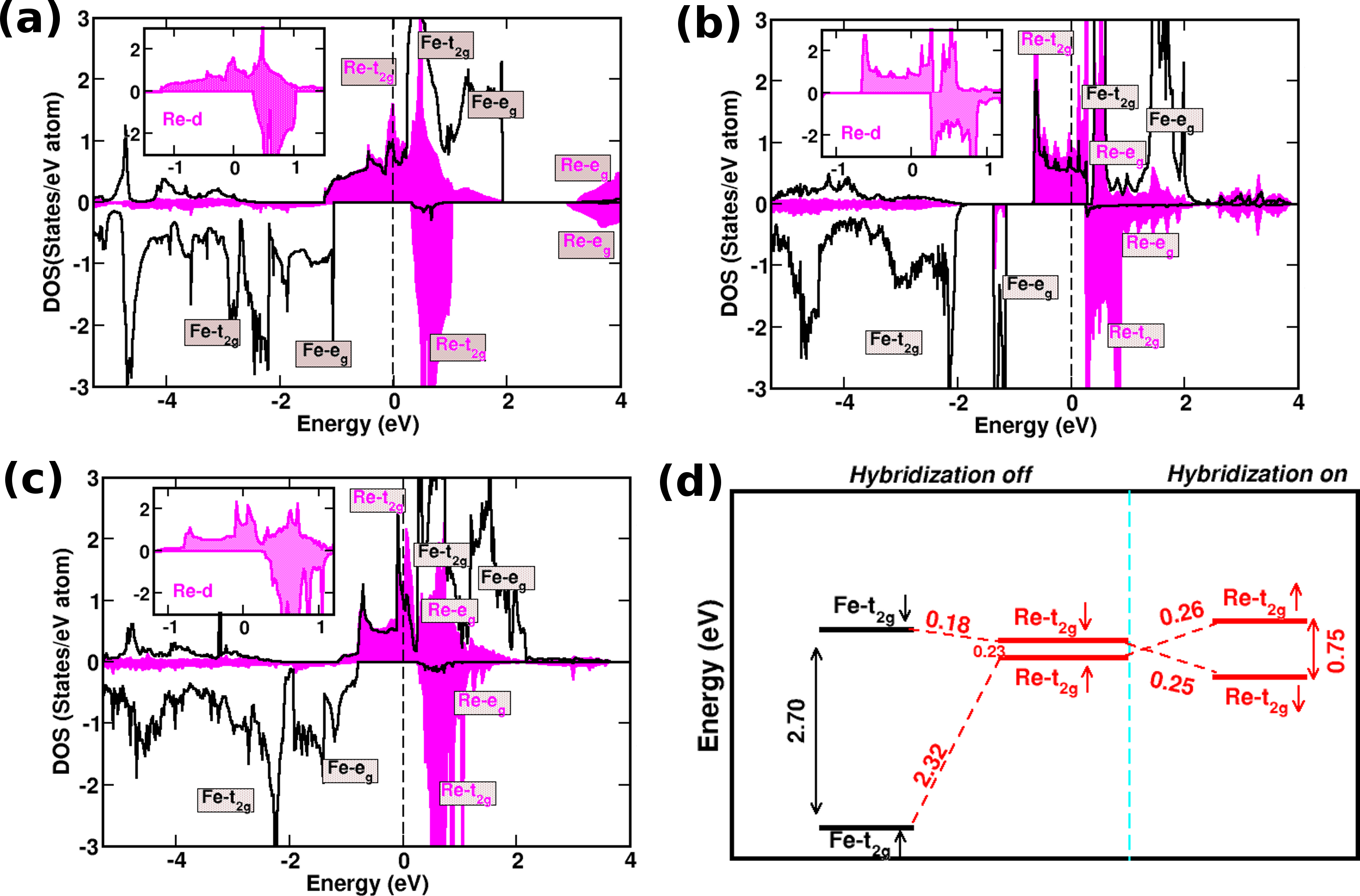}
\caption{(Color online) Density of states, projected onto Fe $d$ (solid line) and Re $d$ states (shaded area), in bulk BFRO [panel (a)],
and in
heterostructures along (111) [panel (b)] and (001) [panel (c)]. The zero of the energy is set at E$_F$. Panel insets
show the zoomed DOS close to E$_F$.  Panel (d) depicts the energy levels
of Fe $t_{2g}$ and Re $t_{2g}$ in absence of the Fe-Re hybridization, and the renormalized Re $t_{2g}$ energy positions after 
hybridization with Fe $t_{2g}$ states. See text for details.}
\end{figure}

The GGA density of states (DOS), projected onto Fe $d$ and Re $d$ states in bulk BFRO as well as (111) and (001) heterostructures 
are shown in Fig. 2. The $d$ states of Fe and Re are spin split as well as split by crystal field into nearly
degenerate sets of $t_{2g}$'s and $e_g$'s.
Considering the bulk DOS, Fe $d$ states are found to be completely filled in the spin-down channel, separated by a gap of $\approx$ 1.2 eV 
from the empty Re $t_{2g}$ states. In the spin-up channel, the Re $t_{2g}$ states strongly hybridized with Fe $t_{2g}$ states cross the Fermi level 
(E$_F$). This makes the system half-metallic with finite density of states at E$_F$ in 
spin-up channel and zero in the spin-down channel, in agreement with literature.\cite{PhysRevLett.98.017204,Wu} 
{\it Remarkably, we find this continues to be the case even for (111) and (001) heterostructures 
with minimum thicknesses possible for BFRO in (111) and (001) directions.} The inclusion of SOC is found to preserve the half-metallic character
for both bulk BFRO and the heterostructures.\cite{note-so} The large onsite energy difference between Ti$^{4+}$ states and that 
of Fe/Re $d$ states, makes the hybridization between the bilayer and the matrix of BTO negligibly small; the electronic structure 
close to E$_F$ is thus dominated by Fe and Re $d$ states. The bandwidths of the states are however significantly narrower than in bulk BFRO, 
especially in case of (111) heterostructure. Interestingly, we find the Re $t_{2g}$-Fe $t_{2g}$ dominated states close to E$_F$ in heterostructures 
show signature of 1D features characterized by van Hove singularities (see insets in Figs. 2(b) and 2(c)), signaling dimensional reduction.

The electronic structure of BFRO appears similar to the highly discussed DP Sr$_{2}$FeMoO$_6$ \cite{sfmo} or 
Cr-based 3d-5d DPs,\cite{3d-5d} suggesting that the hybridization-driven mechanism is operative for magnetism of BFRO, as in these
other cases. In order to probe this further, we calculate the energy 
level positions of Fe $t_{2g}$'s and Re $t_{2g}$'s in bulk BFRO employing NMTO downfolding. 
In absence of hybridization between Fe $t_{2g}$'s and Re $t_{2g}$'s, obtained considering a Fe  $t_{2g}$- Re $t_{2g}$ active
basis set, the energy levels of Re $t_{2g}$ states appear in between the spin split Fe $t_{2g}$ states [cf Fig. 2(d)]. 
Considering the Re $t_{2g}$ only active basis set, which takes into account hybridization effect between Fe $t_{2g}$ and Re $t_{2g}$ states,
it is found that the hybridization between Fe $t_{2g}$ and Re $t_{2g}$ states, pushes up (down) the renormalized Re $t_{2g}$ up (down) 
spin states. This  leads to a renormalized spin splitting at Re site directed oppositely to that of Fe, enforcing ferromagnetic alignment 
of Fe spins, induced solely by hybridization effect. This hybridization driven mechanism works whenever Re atoms have appreciable number of Fe neighbors 
connected through sharing O, as in case of (111) and (001) heterostructures.  

\begin{table}
\caption{The spin (orbital) magnetic moment in $\mu_{B}$, calculated within GGA+SO approximation, at Fe and Re sites for bulk Ba$_{2}$FeReO$_{6}$,
and (111) and (001) bilayers of Ba$_{2}$FeReO$_{6}$.}\label{tab:momentggaso}
%\resizebox{\textwidth}{}{%
\begin{tabular}{cc|c|c}
\hline \hline
\multicolumn{2}{c}{Bulk } & \multicolumn{1}{c}{(111)} & \multicolumn{1}{c}{(001)} \\ \hline
  Ions    &  Spin (Orbital) & Spin (Orbital) &  Spin (Orbital) \\ 
\hline
  Re      & 0.78 (-0.13) &   0.79 (-0.13) & 0.73 (-0.14)  \\
  Fe      &-3.73 (-0.09) &  -3.68 (-0.08) & -3.71 (-0.09) \\
\hline \hline
\end{tabular}
%}
\end{table}

The calculated spin and orbital magnetic moments at Re and Fe sites are listed in Table I. For bulk BFRO, these are in reasonable 
agreement with that predicted from neutron powder diffraction.\cite{PhysRevLett.98.017204} 
\textcolor{black}{We note here the moment at Re ion is somewhat different between our calculation and that reported by Wu.\cite{Wu}
This deviation results from use of different basis set, different exchange-correlation functional and most importantly neglect of 
spin-orbit coupling in the calculation of Wu.\cite{Wu}} 
The moments are found to be rather similar 
between the bulk BFRO and the heterostructures, 
reconfirming that the basic electronic structure of BFRO remains intact in the heterostructure, except for the reduction in 
dimensionality.
\textcolor{black}{To check the effect of correlation beyond GGA, we carried out GGA+U+SOC calculations, with a choice of U$_{Fe}$ = 4 eV,  
U$_{Re}$ = 2 eV and J$_{H}$ = 0.8 eV. The qualitative picture is found to survive, specially the half-metallic character, with larger moments 
on Fe and Re due to slight reduction in bandwidths. The  spin(orbital) moments 
are found to be  between 
-4.1 and -4.2 $\mu_{B}$ (-0.04  and -0.06 $\mu_{B}$) at Fe site and 
-1.1 and -1.3  $\mu_{B}$ (-0.10 and -0.12 $\mu_{B}$) at Re site.} 

\begin{figure}
\includegraphics[width=8.6cm,keepaspectratio]{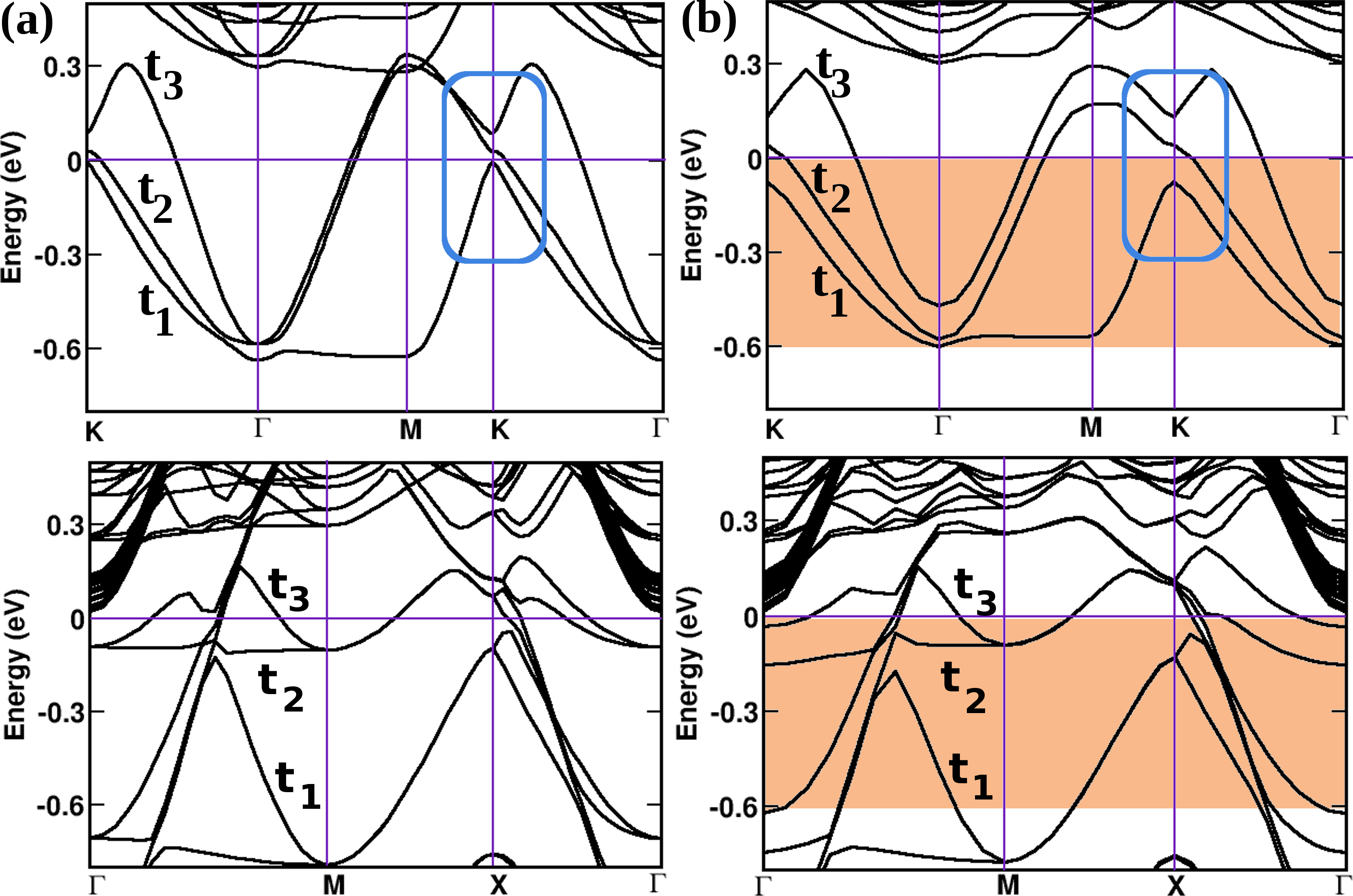}
\caption{(a) The GGA (left panels) and GGA+SO (right panels) band structures of (111) (top panels) and (001) (bottom panels) heterostructures 
plotted along the high symmetry directions. For (111) heterostructure, the chosen high symmetry directions are K = (0.3333, 0.3333, 0), 
to $\Gamma$ = (0, 0, 0) to M = (0.5, 0.0, 0)= to K to $\Gamma$. For (001) heterostructure, the chosen high symmetry directions are 
$\Gamma$ to M =  (0.5, 0.5, 0.0) to X =(0.0, 0.5, 0.0) to $\Gamma$. Zero of the energy is set at respective E$_F$'s.
Blue rectangular area highlights the band topology at high symmetry K point in (111) heterostructure.}
\label{fig:bandtopology}
\end{figure}

To appreciate the effect of SOC in the band structure in heterostructure geometry, which may have important bearing on the
topological nature of the bands of the bilayer, we  show in Fig. 3, the band structure of both the heterostructures 
computed within GGA+SO, in comparison to that of GGA.  In case of (111) heterostructure, three isolated Re $t_{2g}$ bands cross E$_F$ 
having admixture from Fe. The band structure of (001) heterostructure, on the other hand, is far more complicated with
both Fe $d$ and Ti $d$ states crossing the manifold of Re $t_{2g}$ bands. While the effect of SOC is found to be not 
very significant for (001) case, in case of (111) heterostructure the effect is found to be significant.
We find the (111) GGA band structure at the K-point shows near degeneracy of the three bands. This follows the 
small trigonal distortion at Re site, leading to only tiny trigonal splitting. Switching on the SOC coupling pushes the 
 t$_1$ and t$_3$ bands away from each other at K point, though the whole band structure 
remains as metal with highly dispersive bands crossing E$_F$. Since the (111) bilayer band structure consists of isolated 
bands near E$_F$ a Chern number of each of these bands can be defined. The three isolated, spin-polarized, and dominantly 
Re bands near the Fermi level can be modeled using a tight-binding model for t$_{2g}$ orbitals
on the 2D triangular lattice formed by Re atoms on the (111) face. Computing the Berry curvature
of the resulting tight-binding bands (see SM), we find that, in increasing
order of energy at the $\Gamma$-point, the bands have Chern numbers $-2,2,0$. Thus, the combination of SOC and low
dimensionality, drives topological half-metallic bands with large Chern number. \textcolor{black}{Though the t$_1$ and t$_2$ 
bands are found to have nontrivial Chern numbers, the solution turns out to be metallic rather than insulating. This happens 
as the crystal field splitting is not large enough, needed to push the t$_1$ and t$_2$ bands sufficiently far from each other. 
The (111) bilayer thus can be
a quantum anomalous Hall insulator, if t$_1$ and t$_2$ bands can be prevented from spanning a common energy window. This 
may be possible through strain engineering, which will be a topic of future study.}

The presence of van Hove singularities in the low energy electronic structure of (111) and (001) bilayers
strongly suggests reduction of dimensionality of the half-metallic conducting electrons compared to bulk BFRO. 
From the plot of the charge density, as shown in Figs 4(a) and 4(b) for (111) heterostructure, and Figs 4(c) and 
4(d) for (001) heterostructure, we find the conduction 
charges to be entirely localized in the bilayer of BFRO, with little contribution in BTO block. The conducting charge 
confined in the BFRO bilayer are distributed within the Re and Fe sites, connected via oxygens forming a highly mobile and spin-polarized
2DEG. In order to have quantitative estimates of the confinement, the layer averaged charge density is computed and plotted 
as a function of position along the growth direction. The results are shown in Fig 4(e). For 
nonpolar (001) interfaces, the concentration profile is symmetric, with a spread of about $14$\AA. Changing the growth direction to (111) with 
creation of polar interfaces makes the concentration profile asymmetric with a drastic reduction in spread
to only about $7$\AA, which is of the order of unit cell thickness of BFRO. We contrast this with the confinement of the
2DEG formed at the (001) LAO/STO interface, keeping in mind that the origin
of confinement is different in the two cases. 
The infrared ellipsometry data \cite{prl} shows the vertical concentration profile of the electrons at the LAO/STO interface 
has a strong asymmetric shape with a initial decay over $2$ nm and a pronounced tail that extends to $\sim 11$ nm,
an order of magnitude larger than in BTO/BFRO/BTO.

\begin{figure}
\includegraphics[width=7.8cm,keepaspectratio]{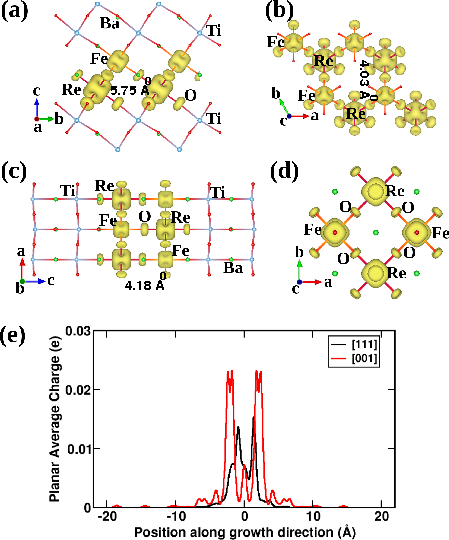}
\caption{\textcolor{black}{Out-of-plane and inplane projection of the conducting charge density in (111) [panels (a) and (b), respectively] and (001) 
heterostructure geometries [panels (c) and (d), respectively]. The isosurface values are chosen as $0.023{\rm e}^{-}/$\AA$^3$. The charge 
densities are calculated from contributions within an energy window [$E_F-0.5 eV,E_F$]. Various atomic distances have been marked.
Panel (e): Planar averaged conducting charge density versus distance along the growth direction for (111) and (001) heterostructures. The
distances are measured from the centre of the unit cell, which is at the mid point of the bilayer.}}
\end{figure}

We compute effective masses corresponding to the three $t_{2g}$ bands at $\Gamma$ point showing parabolic
or near parabolic dispersion. The bands falling in this region (shaded) are marked in Fig. 3. Effective mass tensor 
$m^{-1}_{ij} = \frac{1}{\hbar} \frac{\partial^2 E}{\partial k_{i} \partial k_{j}}$ is computed using finite difference method in unit of 
rest mass $m_{0}$. The inplane and out-of-plane effective masses for (111) and (001)  
heterostructures are tabulated in Table II, which further brings out the highly confined 2D mobile character of the carriers.

\begin{table}[t]
\caption{Inplane and out-of-plane effective masses of the conducting bands for (111) and (001) heterostructures, calculated at $\Gamma$ point.}
\begin{tabular}{c|c|c|c|c|c|c|c}
\hline \hline
             & Band\# &  $m_{\parallel}^{eff}$ & $m_{\perp}^{eff}$ &  & Band\# &  $m_{\parallel}^{eff}$ & $m_{\perp}^{eff}$\\
  (111)      & 3. &  1.2 $m_{0}$ &  22 $m_{0}$ & (001) & 3. &  2.3 $m_{0}$ &  73 $m_{0}$ \\\
             & 2. &  0.8 $m_{0}$ &  840 $m_{0}$ & & 2. &  1.0 $m_{0}$ &  73 $m_{0}$\\
             & 1. &  3.6 $m_{0}$ &  61 $m_{0}$ & & 1. &  7.8 $m_{0}$ &  5.6 $m_{0}$\\
\hline \hline
\end{tabular}
\end{table}

To conclude, our DFT study establishes that BTO/BFRO/BTO heterostructure geometry generates spin polarized 2D electron gas at the interface 
confined solely to the BFRO bilayer, which retains the half-metallic character of bulk BFRO. The confinement is driven by suppression
of hybridization, a mechanism distinct from the polar catastrophe driven confinement of the 2DEG realized at the (001) LAO/STO or 
GTO/STO or SmTO/STO interfaces. For the (111) orientation, the confinement of the 2DEG is greatly aided by polar fields at the interface,
leading to confinement width of the order of unit cell thickness and an order magnitude smaller than that reported for LAO/STO.
This highly confined 2D electron gas which is half-metallic in nature, with a moment $\approx$ 3.0 $\mu_B$/unit cell, should be promising for 
applications in ultra-thin spintronic devices. Bulk BFRO has a room temperature magnetic transition 
temperature of 304 K and a significant magnetocrystalline anisotropy $\approx 1$-$2$ meV.\cite{Plumb2013} The proposed 
half-metallic quantum well structure is thus expected to be operative at a reasonably high temperature and stable against
magnetic spiral instabilities due to Rashba SOC. We further find that the interface 
bands of (111) structure have topological character, with the potential to realize the quantum anomalous Hall effect.  
With insulating nature of the BTO layers, this heterostructure also naturally permits top or bottom
gating to control the transport in the 2DEG, making it suitable for hybrid spintronic and electronic devices.
Our computational study provides strong motivation for experimental investigations of half-metallic 2DEGs in oxide quantum wells.

S.B. and T.S.D thank Department of Science and Technology, India for the support through Thematic Unit 
of Excellence. AP was supported by NSERC (Canada).

\vskip 2in
$\ast$ Presently at University of Duisburg-Essen, Duisburg. \\
$\dagger$ Email: t.sahadasgupta@gmail.com

\end{document}